\documentclass[aip,apl,reprint]{revtex4-2}
\usepackage{amsmath,amssymb}
\usepackage{graphicx}
\usepackage{booktabs}
\usepackage{tabularx}

\begin{document}

\title{Variance-Driven Mean Temperature Reduction in Nonuniformly Heated Radiative--Conductive Systems}
\author{Juntao Lu}
\affiliation{Zhenhai High School of Zhejiang, Ningbo 315200, China}
\author{Zihan Zhang}
\affiliation{Zhenhai High School of Zhejiang, Ningbo 315200, China}
\author{Yongjian Xiong}
\affiliation{School of Physical Science and Technology, Ningbo University, Ningbo 315211, China}
\author{Jie Fu}
\email{fujie@nbu.edu.cn}
\affiliation{Institute of High-Pressure Physics, School of Physical Science and Technology, Ningbo University, Ningbo 315211, China}
\date{\today}

\begin{abstract}
Radiative-conductive systems are intrinsically nonlinear due to the quartic temperature dependence of thermal radiation. 
Under fixed total heating power, convexity arguments imply that nonuniform temperature distributions radiate 
more efficiently and therefore exhibit a lower mean temperature than their isothermal counterparts. 
However, this conclusion remains qualitative, and an explicit quantitative relation between temperature heterogeneity 
and mean temperature reduction has been lacking. 
Here we derive a variance-based analytical expression linking the area-averaged temperature to the corresponding isothermal equilibrium temperature in a nonuniformly heated radiative--conductive system. By integrating the governing equation and performing a systematic second-order expansion about the ambient temperature, we show that the decrease of the mean temperature relative to the isothermal equilibrium value is linearly proportional to the temperature variance, with a proportionality coefficient set solely by the ambient temperature. This result transforms the convexity-based inequality into a quantitative statistical relation within the perturbative regime and provides a physically transparent framework for describing nonlinear radiative averaging in thermally heterogeneous systems.
\end{abstract}

\maketitle

\section{INTRODUCTION}
Radiative heat transfer inherently introduces strong nonlinearity into thermal systems due to the quartic temperature dependence of the Stefan-Boltzmann law.\cite{Asllanaj2003,Saldanha1997} 
When radiation is coupled with conduction, nonuniform heating gives rise to spatially varying temperature fields, and the resulting radiative loss becomes highly sensitive to temperature heterogeneity.\cite{Venkatraman1997Fluctuation}
Such nonisothermal radiative systems are not only theoretically nonlinear but also practically significant, as spatial temperature control has recently been shown to enable directional and enhanced thermal emission.\cite{Herz2025,Krebs1996Fluctuations}

Under fixed total heating power, a nonuniform temperature distribution radiates more efficiently than an isothermal one because of the convex
nature of the $T^4$ dependence. This implies that the area-averaged temperature of a nonisothermal system must be lower than that of the
corresponding isothermal state satisfying the same global power balance. Existing studies on radiative-conductive coupling mainly focus on
the existence, uniqueness, or numerical solution of nonlinear governing
equations.\cite{Asllanaj2003,Thompson2004,Ghattassi2019,Talukdar2002CDM,GuentherLee1998} While this qualitative effect is well recognized in nonlinear radiative
systems,\cite{Zhang2023ApJ} a clear and explicit quantitative relation between the mean temperature reduction, which is truly experimentally accessible, and the degree of temperature nonuniformity has not been established.

In this work, we analyze a thin circular disk subject to localized volumetric heating and radiative exchange. By integrating the nonlinear governing equation and performing a systematic second-order expansion about the ambient temperature, we derive an analytical expression linking the reduction of area-averaged temperature to the temperature variance. The resulting relation provides an explicit description of nonlinear radiative averaging induced purely by spatial thermal heterogeneity, transforming the convexity-based inequality into a quantitative statistical relation within the perturbative regime and offering a physically transparent framework.

\section{Model description and governing equation}
\begin{figure}[htbp]
\centering
\includegraphics[width=\columnwidth]{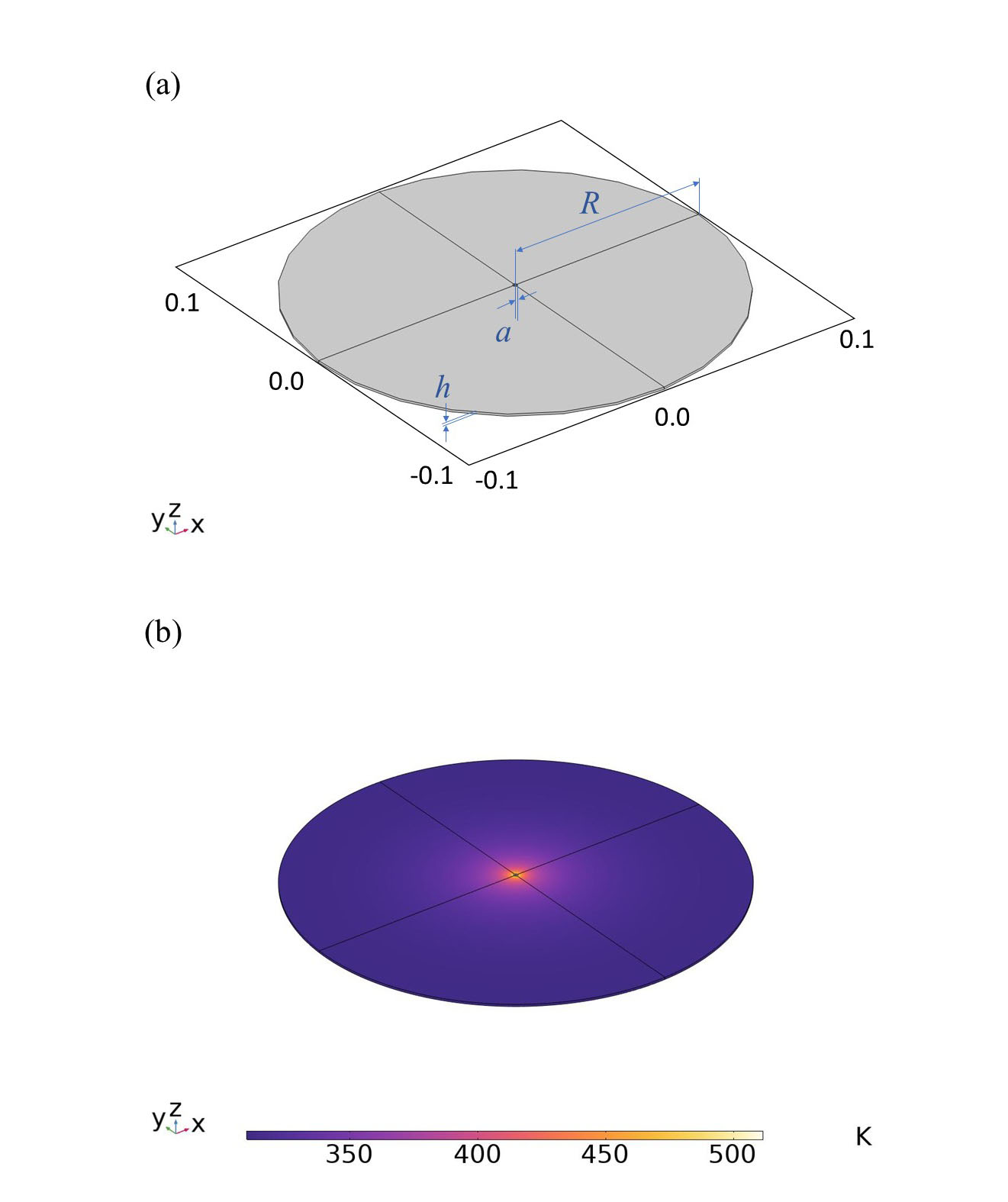}
\caption{(a) Geometry of the radiative--conductive thin disk model. (b) Representative axisymmetric temperature distribution under localized heating at center.}
\label{fig:1}
\end{figure}

We consider a thin circular disk of radius $R$ and thickness $h$, made of an isotropic material with thermal conductivity $k$. 
The disk is subjected to an axisymmetric volumetric heat generation $Q(r)$ (W$\cdot$m$^{-3}$), which is localized within a central region $0 \le r \le a$, as shown in Fig. ~\ref{fig:1}, such that
\[
Q(r) =
\begin{cases}
Q_0, & 0 \le r \le a, \\
0, & a < r \le R.
\end{cases}
\]

\begin{table}[htbp]
\centering
\begin{tabular}{ll}
\toprule
Symbol     & Description \\
\midrule
$R$        & Radius of disk \\
$h$        & Thickness of disk \\
$k$        & Thermal conductivity \\
$\epsilon$  & Emissivity \\
$\sigma$    & Stefan-Boltzmann constant \\
$Q_0$     & Volumetric heat source density at center \\
$a$        & Radius of heat source \\
$T_{\mathrm a}$        & Ambient temperature \\
\bottomrule
\end{tabular}
\caption{Nomenclature}
\end{table}
The disk exchanges thermal radiation with a large isothermal surrounding at temperature $T_{\mathrm a}$ (taken as ambient temperature), characterized by a hemispherical emissivity $\varepsilon$. Radiation is assumed to occur at the upper surface only. The edge and bottom are assumed adiabatic.
Convection is neglected in order to isolate the conductive--radiative coupling.
Starting from the Fourier's law,\cite{Ozisik1993} volumetric heat flux is:
\begin{equation}
\mathbf{q} =  - k \nabla T.
\label{eq:fourier}
\end{equation}
We adopt an axisymmetry and thin-plate approximation $h \ll R$, 
such that the temperature is nearly uniform across the thickness 
and can be represented by the mid-plane field $T(r)$.\cite{LeeChoi2021ThinPlate}
Integrating the volumetric heat flux across the thickness $h$ yields the effective in-plane surface heat flux
\begin{equation}
\mathbf{q}\mathrm{_r} =  - k h \frac{\operatorname{d}T}{\operatorname{d}r} \hat{r}.
\label{eq:inplane}
\end{equation}
Applying steady-state energy conservation \cite{Whitaker1977} and Stefan-Boltzmann law yields:
\begin{equation}
\nabla \cdot \mathbf{q}\mathrm{_r} +  \varepsilon \sigma (T^4 - T_{\mathrm a}^4) = hQ(r),
\label{eq:energy1}
\end{equation}
where $hQ(r)$ represents the heat input per unit surface area obtained by integrating the volumetric source across the thickness. All terms in Eq.~\eqref{eq:energy1} therefore have units of heat flux per unit area (W m$^{-2}$). Therefore,

\begin{equation}
\frac{1}{r} \frac{\operatorname{d}}{\operatorname{d}r}
\left(
r \frac{\operatorname{d}T}{\operatorname{d}r}
\right)
- \frac{\varepsilon \sigma}{k h}
\left( T^4 - T_{\mathrm a}^4 \right)
+ \frac{Q(r)}{k}
= 0,
\qquad 0 \leqslant  r \leqslant  R.
\label{eq:energy2}
\end{equation}
Regularity at the disk center requires
\begin{equation}
\left.
\frac{\operatorname{d}T}{\operatorname{d}r}
\right|_{r=0}
= 0,
\label{eq:bc1}
\end{equation}
and an adiabatic outer boundary condition is imposed as
\begin{equation}
\left.
\frac{\operatorname{d}T}{\operatorname{d}r}
\right|_{r=R}
= 0.
\label{eq:bc2}
\end{equation}
For convenience, we introduce the radiative coupling parameter
\begin{equation}
\alpha = \frac{\varepsilon \sigma}{k h}.
\label{eq:alpha}
\end{equation}
Eventually, the governing equation can be written compactly as
\begin{equation}
\frac{1}{r} \frac{\operatorname{d}}{\operatorname{d}r}
\left(
r \frac{\operatorname{d}T}{\operatorname{d}r}
\right)
- \alpha \left( T^4 - T_{\mathrm a}^4 \right)
+ \frac{Q(r)}{k}
= 0.
\label{eq:ge}
\end{equation}

\section{Validation of the thin-plate reduction}
To assess the validity of the thin-plate approximation introduced in Sec.~II, the full three-dimensional (3D) heat conduction problem with surface radiation boundary conditions is solved using finite-element simulations in COMSOL Multiphysics. The model consists of a solid disk of thickness $h$ and radius $R$, with radiative heat exchange imposed at the upper surface, while the bottom surface and outer edge are treated as adiabatic boundaries. The volumetric heat source is implemented according to the distribution defined in Sec.~II. Considering a ceramic disk,\cite{Ballan2022TiC} specific parameters used in the calculations are listed in Table\label{tab:parameters}

\begin{table}[htbp]
\centering
\begin{tabular}{ll}
\toprule
Parameter & {Value} \\
\midrule
$R\;(\mathrm{m})$                & 0.1 \\
$h\;(\mathrm{m})$                & 0.001 \\
$k\;(\mathrm{W\,m^{-1}\,K^{-1}})$ & 10 \\
$\epsilon$                       & 0.8 \\
$Q_0\;(\mathrm{W\,m^{-3}})$      & {$10^{9}$} \\
$a\;(\mathrm{m})$                & 0.001 \\
$T_{\mathrm a}\;(\mathrm{K})$              & 300 \\
\bottomrule
\end{tabular}
\caption{Geometrical dimensions, material properties, and source parameters used for validating the thin-plate reduction.}
\label{tab:parameters}
\end{table}

The steady-state temperature field obtained from the 3D simulation is compared with the solution of the reduced two-dimensional (2D) governing equation derived from the thickness-integrated energy balance, as Eq.~\eqref{eq:ge}. This equation is solved using a collocation-based boundary value solver (bvp4c) in Matlab under the boundary conditions defined in Eqs.~\eqref{eq:bc1} and ~\eqref{eq:bc2}. 

\begin{figure}[htbp]
\centering
\includegraphics[width=\columnwidth]{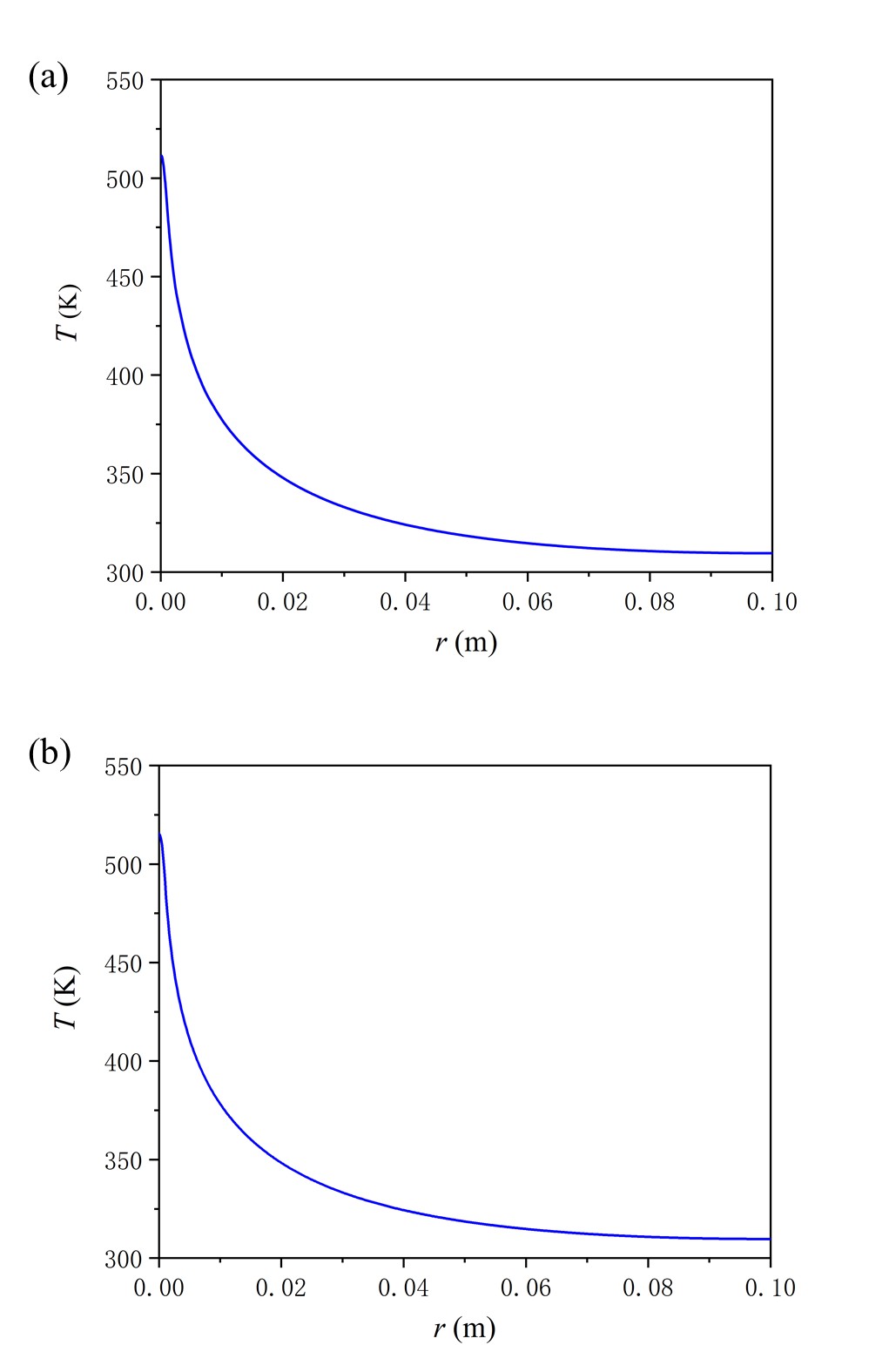}
\caption{Comparison between the full 3D finite-element simulation and the reduced 2D model: (a) mid-plane temperature extracted from the 3D simulation; (b) numerical solution of the reduced governing equation.}
\label{fig:2}
\end{figure}

Eventually, the mid-plane temperature distribution $T(r)$ and the maximum temperature rise are examined, as shown in Fig.~\ref{fig:2}. The 2D solution is found to be virtually indistinguishable from the mid-plane temperature extracted from the full 3D simulation. The relative deviation in peak temperature remains below 0.84\% for the parameters considered. These results validate the thin-plate reduction and demonstrate that the 2D model accurately captures the steady-state thermal behavior under conductive--radiative coupling.

\section{Analytical relation between mean temperature and temperature variance}
We begin by defining the area-averaged temperature of the nonisothermal surface,
\begin{equation}
\bar{T}
=
\frac{1}{A}
\int_A T\, \operatorname{d}A
=
\frac{2}{R^2} \int rT(r)\, \operatorname{d}r,
\qquad
A=\pi R^2.
\end{equation}

For an isothermal surface with uniform temperature $T_{\mathrm{iso}}$, steady state requires that the total
radiative heat loss equals the total input power. This condition gives
\begin{equation}
T_{\mathrm{iso}}
=
\left(
T_\mathrm{a}^4
+
\frac{a^2 h Q_0}{\sigma \epsilon R^2}
\right)^{1/4}.
\end{equation}

\begin{figure}[htbp]
\centering
\includegraphics[width=\columnwidth]{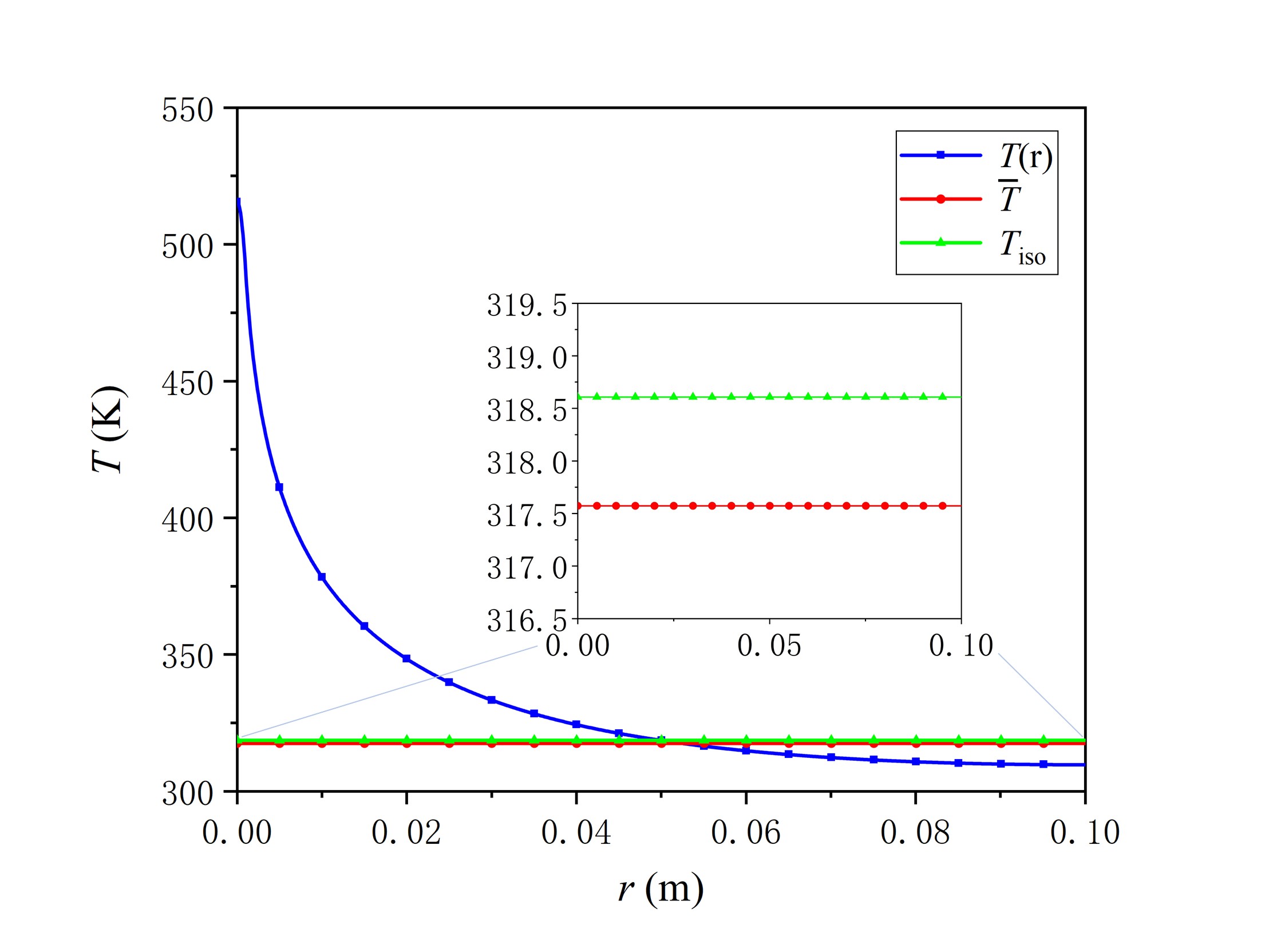}
\caption{
Radial temperature profile $T(r)$ under localized heating. 
The horizontal lines denote the area-averaged temperature $\bar T$ and the corresponding isothermal equilibrium temperature $T_{\rm iso}$. The inset is the zoomed-in comparison between $\bar T$ and $T_{\rm iso}$, demonstrating the reduction of the mean temperature in the nonisothermal case.
}
\label{fig:3}
\end{figure}

Figure~\ref{fig:3} shows the numerical comparison between $\bar T$ and $T_{\rm iso}$ under a radial temperature profile of $T(r)$ by numerically solving the reduced 2D governing Eq.~\eqref{eq:ge}. The average temperature of the nonisothermal disk $\bar T$ is lower than the isothermal one $T_{\rm iso}$, consistent with the convexity of the radiative heat-loss term.

Instead, our aim is to find the quantitative difference between $T_{\mathrm{iso}}$ and $\bar{T}$ analytically. 
We integrate the steady-state governing equation over the disk area $A$:

\[
\int_A \left[
  \frac{1}{r} \frac{\operatorname{d}}{\operatorname{d}r}\left(r \frac{\operatorname{d}T}{\operatorname{d}r}\right)
  - \frac{\varepsilon \sigma}{k h}\left( T^4 - T_{\text{a}}^4 \right)
  + \frac{Q(r)}{k}
\right] \operatorname{d}A = 0.
\]

Using the adiabatic boundary condition on the rim, the divergence term vanishes and yields the global power balance

\begin{equation}
\varepsilon \sigma \int_A \left( T^4 - T_{\mathrm a}^4 \right)\, \operatorname{d}A
=
\int_A h\,Q(r)\, \operatorname{d}A
\equiv P_{\text{in}},
\end{equation}
where $P_{\mathrm{in}}$ is the heat input power. Defining the area average $\langle f \rangle = \frac{1}{A}\int_A f\,dA$,
we obtain

\begin{equation}
\varepsilon \sigma A \left( \langle T^4 \rangle - T_{\mathrm a}^4 \right)
=
P_{\text{in}} .
\end{equation}

The isothermal temperature $T_{\mathrm{iso}}$ satisfies the same
power balance,

\begin{equation}
\varepsilon \sigma A \left( T_{\mathrm{iso}}^4 - T_{\mathrm a}^4 \right)
=
P_{\text{in}} .
\end{equation}

Comparing the two expressions gives the key identity

\begin{equation}
\langle T^4 \rangle = T_{\mathrm{iso}}^4 .
\end{equation}

This identity follows solely from the global radiative power balance and is independent of the specific geometry, heat-source distribution, or thermal conductivity. It reflects a structural consequence of the quartic nonlinearity of radiative heat loss rather than of conductive transport.\cite{Kim2013EffectiveTemp}

Let $T = T_{\mathrm a} + \theta$, with $|\theta| \ll T_{\mathrm a}$.
Expanding to second order,

\begin{equation}
(T_{\mathrm a} + \theta)^4
=
T_{\mathrm a}^4
+ 4 T_{\mathrm a}^3 \theta
+ 6 T_{\mathrm a}^2 \theta^2
+ O(\theta^3).
\label{eq:Taylor}
\end{equation}

Taking area averages,

\begin{equation}
\langle T^4 \rangle
=
T_{\mathrm a}^4
+ 4 T_{\mathrm a}^3 \langle \theta \rangle
+ 6 T_{\mathrm a}^2 \langle \theta^2 \rangle
+ O(\theta^3).
\label{eq:avg-t}
\end{equation}

Similarly, writing
$T_{\mathrm{iso}} = T_{\mathrm a} + \theta_{\mathrm{iso}}$ and expanding,

\begin{equation}
T_{\mathrm{iso}}^4
=
T_{\mathrm a}^4
+ 4 T_{\mathrm a}^3 \theta_{\mathrm{iso}}
+ 6 T_{\mathrm a}^2 \theta_{\mathrm{iso}}^2
+ O(\theta^3).
\label{eq:avg-iso}
\end{equation}

Using $\langle T^4\rangle = T_{\mathrm{iso}}^4$ and substituting 
Eqs.~\eqref{eq:avg-t} and ~\eqref{eq:avg-iso}, we obtain

\begin{equation}
T_{\mathrm a}^4 + 4T_{\mathrm a}^3 \langle \theta \rangle 
+ 6T_{\mathrm a}^2 \langle \theta^2 \rangle
=
T_{\mathrm a}^4 + 4T_{\mathrm a}^3 \theta_{\mathrm{iso}}
+ 6T_{\mathrm a}^2 \theta_{\mathrm{iso}}^2
+ O(\theta^3).
\end{equation}

Cancelling $T_{\mathrm a}^4$ and dividing both sides by $4T_a^3$ yields

\begin{equation}
\langle \theta \rangle
+
\frac{3}{2T_{\mathrm a}}\langle \theta^2 \rangle
=
\theta_{\mathrm{iso}}
+
\frac{3}{2T_{\mathrm a}}\theta_{\mathrm{iso}}^2
+ O(\theta^3).
\end{equation}

Thus,
\begin{equation}
\langle \theta \rangle = \theta_{\mathrm{iso}} + O(\theta^2),
\qquad
\langle \theta \rangle^2 = \theta_{\mathrm{iso}}^2 + O(\theta^3).
\end{equation}

Introducing the variance
\[
\operatorname{Var}(\theta)
=
\langle \theta^2 \rangle
-
\langle \theta \rangle^2,
\]
we write
\[
\langle \theta^2 \rangle
=
\operatorname{Var}(\theta)
+
\langle \theta \rangle^2.
\]

Substituting this expression and noting that 
$\langle \theta \rangle^2$ and $\theta_{\mathrm{iso}}^2$
are second-order small quantities, we retain only 
terms up to $O(\theta^2)$, which gives

\begin{equation}
\langle \theta \rangle
+
\frac{3}{2T_{\mathrm a}}\operatorname{Var}(\theta)
=
\theta_{\mathrm{iso}}
+ O(\theta^3).
\end{equation}

Therefore,

\begin{equation}
\langle \theta \rangle
=
\theta_{\mathrm{iso}}
-
\frac{3}{2T_{\mathrm a}}\operatorname{Var}(\theta)
+ O(\theta^3).
\end{equation}

Recalling that $\bar T = T_{\mathrm a} + \langle \theta \rangle$ and 
$T_{\mathrm{iso}} = T_{\mathrm a} + \theta_{\mathrm{iso}}$, we finally obtain

\begin{equation}
\bar T
=
T_{\mathrm{iso}}
-
\frac{3}{2T_{\mathrm a}}\operatorname{Var}(\theta)
+ O(\theta^3).
\label{eq:23}
\end{equation}

Equation~\eqref{eq:23} shows that the reduction in mean temperature is proportional to the temperature variance with proportionality factor of  $3/(2T_a)$ originating solely from the quartic temperature dependence of radiative heat loss, which reveals that the mean-temperature shift arises as a purely nonlinear statistical effect rather than from conductive details. Figure~4 compares the isothermal equilibrium temperature $T_{\rm iso}$ and the mean temperature $\bar T$ under the radial temperature profile $T(r)$. Two average values of temperature, $\bar T_{\mathrm{num}}$ and $\bar T_{\mathrm{anal}}$, are plotted: one obtained by numerically solving the reduced two-dimensional governing Eq.~\eqref{eq:ge} and the other evaluated from the analytical relation in Eq.~\eqref{eq:23}, respectively. Both results lie below $T_{\rm iso}$, and the analytical prediction $\bar T_{\rm anal}$ is in excellent agreement with the numerical result $\bar T_{\rm num}$, confirming that the mean-temperature reduction is quantitatively governed by the temperature variance to second order in the small-deviation expansion.

\begin{figure}[htbp]
\centering
\includegraphics[width=\columnwidth]{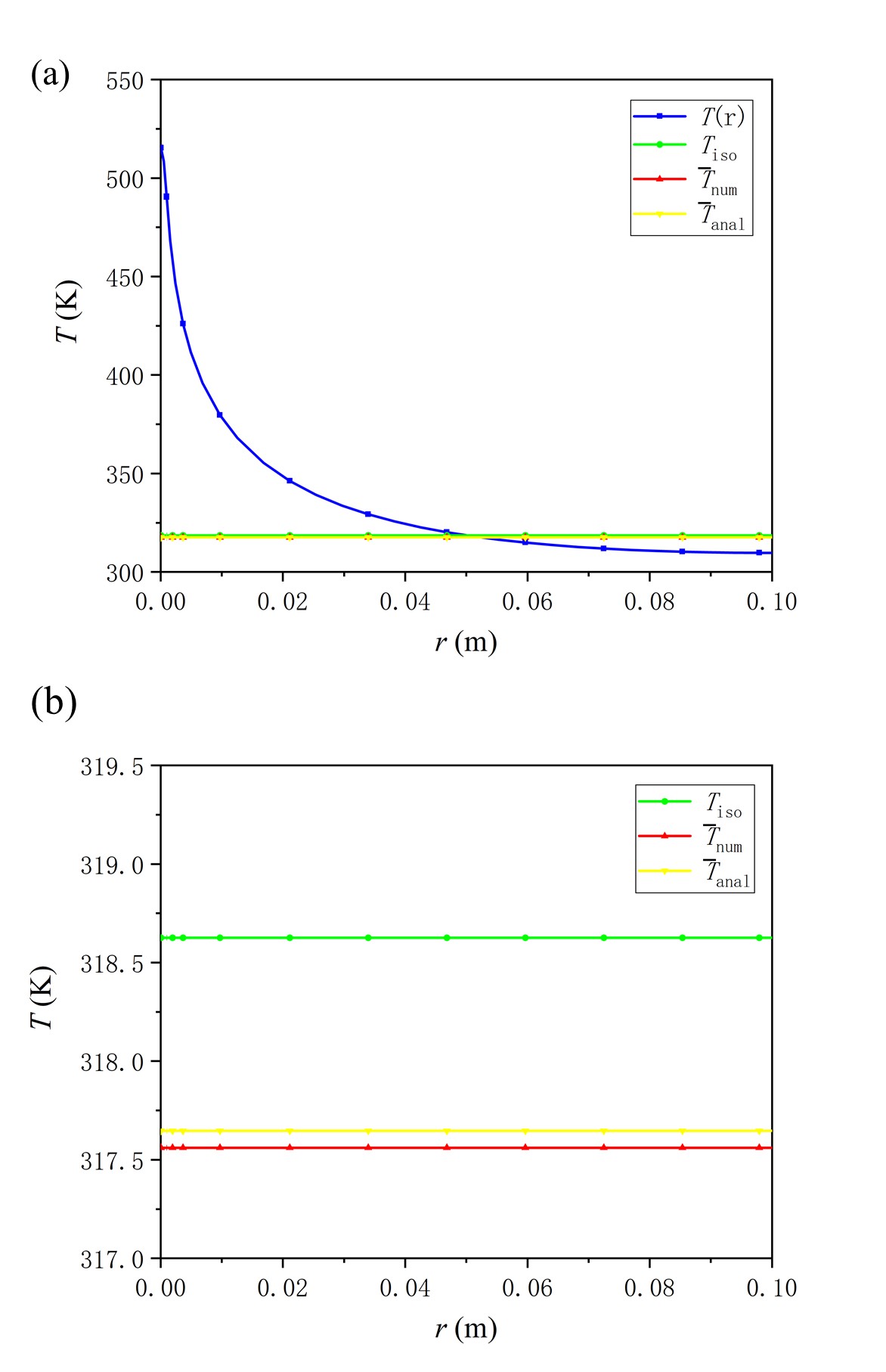}
\caption{
Comparison between the exact mean temperature and the variance-based approximation. 
(a) Radial temperature profile $T(r)$, $T_{\rm iso}$, the numerical mean temperature  $\bar T_{\mathrm{num}}$, and the analytical result $\bar T_{\mathrm{anal}}$. 
(b) Zoomed-in view highlighting the quantitative agreement between $\bar T_{\mathrm{num}}$ and $\bar T_{\mathrm{anal}}$, and both lower than $T_{\rm iso}$.
}
\label{fig:4}
\end{figure}

\section{Validity range of the variance-based relation}

The analytical relation in Eq.~\eqref{eq:23} was derived under the assumption that temperature nonuniformity remain small compared with the ambient temperature, \textit{i.e.}, $|\theta| \ll T_\textrm{a}$. To assess the practical range of validity of this second-order expansion, we vary the heat-source intensity $Q_0$ and quantify the deviation between the numerical mean temperature $\bar T_{\mathrm{num}}$ and the variance-based approximation $\bar T_{\mathrm{anal}}$.

\begin{figure}[htbp]
\centering
\includegraphics[width=\columnwidth]{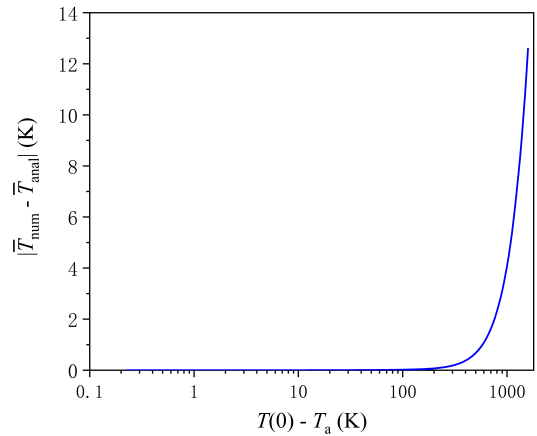}
\caption{Absolute error between the numerical mean temperature 
$\bar T_{\rm num}$ and the second-order analytical approximation 
$\bar T_{\rm anal}$ as a function of the maximum temperature nonuniformity
$T(0)-T_\textrm{a}$. The error is defined as 
$|\bar T_{\rm num}-\bar T_{\rm anal}|$. The logarithmic horizontal axis highlights the broad parameter range over which the second-order approximation remains accurate.}
\label{fig:5}
\end{figure}

Specifically, the maximum temperature nonuniformity $\Delta T_{\max}=T(0)-T_\textrm{a}$ is used as a control parameter. For each value of $Q_0$, we compute the exact mean temperature $\bar T_{\mathrm{num}}$ from the numerical solution of the governing equation and compare it with the analytical prediction $\bar T_{\mathrm{anal}}$ given by Eq.\eqref{eq:23}. Figure~\ref{fig:5} shows the absolute error $|\bar T_{\mathrm{num}}-\bar T_{\mathrm{anal}}|$ as a function of $\Delta T_{\max}$.

The results indicate that the second-order approximation remains highly accurate not only in the strict small-deviation regime, but also when the central temperature rise reaches several hundred kelvin above ambient. This behavior indicates that the relevant perturbative parameter controlling the accuracy of the expansion is not the peak temperature $\Delta T_{\max}$ alone, but the dimensionless variance $\mathrm{Var}(\theta)/T_\textrm{a}^2$. This follows directly from the expansion structure in Eq.~\eqref{eq:Taylor}, where the small parameter naturally appears in the normalized form $\theta/T_a$. Even when the peak temperature rise becomes large, the temperature profile may remain relatively smooth, so that the normalized variance $\mathrm{Var}(\theta)/T_\textrm{a}^2$ stays small. As long as this quantity remains small, the variance-based relation provides an accurate quantitative prediction of the mean-temperature reduction.

\section{Conclusion}

In this work, we established a quantitative statistical relation governing nonlinear radiative--conductive systems under nonuniform heating. Starting from global energy conservation, we identified the exact identity $\langle T^4 \rangle = T_{\mathrm{iso}}^4$, which reveals that radiative heat transfer imposes a nonlinear averaging constraint independent of the detailed heating profile or material parameters.

By performing a second-order expansion about the ambient temperature, we derived the explicit relation \[\bar{T} = T_{\mathrm{iso}} - \frac{3}{2T_\mathrm{a}}\,\mathrm{Var}(\theta) + O(\theta^3).\] It implies that the reduction in mean temperature is directly proportional to the temperature variance. The proportionality coefficient depends solely on the quartic nonlinearity of radiative heat loss, indicating that the cooling advantage of nonuniform heating is a purely nonlinear statistical effect rather than a consequence of conductive redistribution.

Numerical solutions of the reduced governing equation and full three-dimensional finite-element simulations confirm the validity of this variance-based relation over a broad parameter range, extending well beyond the strict asymptotic regime. The accuracy is controlled not by the peak temperature rise itself, but by the dimensionless variance $\mathrm{Var}(\theta)/T_\textrm{a}^2$, highlighting the intrinsic perturbative structure of nonlinear radiative averaging.

The present framework provides a transparent bridge between microscopic temperature heterogeneity and macroscopic cooling performance. More broadly, it suggests that in radiative systems governed by strong nonlinear emission laws, spatial variance emerges as the fundamental perturbative invariant governing nonlinear radiative averaging. This perspective may be extended to other geometries and multiphysics systems in which nonlinear radiative exchange plays a dominant role.

\bibliographystyle{apsrev4-2}
\bibliography{refs} 

@article{Asllanaj2003,
  author  = {F. Asllanaj and G. Jeandel and J. R. Roche and D. Schmitt},
  title   = {Existence and Uniqueness of a Steady State Solution of a Coupled Radiative--Conductive Heat Transfer Problem for a Non-grey Anisotropically and Participating Medium},
  journal = {Transport Theory and Statistical Physics},
  volume  = {32},
  number  = {1},
  pages   = {1--35},
  year    = {2003},
  doi     = {10.1081/TT-120018650}
}

@article{Thompson2004,
  author  = {M. Thompson and C. Segatto and M. T. de Vilhena},
  title   = {Existence Theory for the Solution of a Stationary Nonlinear Conductive--Radiative Heat-Transfer Problem in Three Space Dimensions},
  journal = {Transport Theory and Statistical Physics},
  volume  = {33},
  number  = {5--7},
  pages   = {563--576},
  year    = {2004},
  doi     = {10.1081/TT-200053941}
}

@article{Ghattassi2019,
  author  = {M. Ghattassi and J. R. Roche and D. Schmitt},
  title   = {Analysis of a full discretization scheme for 2D radiative--conductive heat transfer systems},
  journal = {Journal of Computational and Applied Mathematics},
  volume  = {346},
  pages   = {1--17},
  year    = {2019},
  doi     = {10.1016/j.cam.2018.06.028}
}

@article{Zhang2023ApJ,
  author  = {Zhang, Y. and Li, X. and Wang, J. and others},
  title   = {Nonuniform Surface Temperature Effects on Radiative Flux and Effective Temperature},
  journal = {The Astrophysical Journal},
  volume  = {957},
  number  = {1},
  pages   = {20},
  year    = {2023},
  doi     = {10.3847/1538-4357/acf5e3}
}

@article{Ballan2022TiC,
  author  = {Ballan, Michele and Corradetti, Stefano and Manzolaro, Mattia and Meneghetti, Giovanni and Andrighetto, Alberto},
  title   = {Thermal and Structural Characterization of a Titanium Carbide/Carbon Composite for Nuclear Applications},
  journal = {Materials},
  year    = {2022},
  volume  = {15},
  number  = {23},
  pages   = {8358},
  doi     = {10.3390/ma15238358}
}

@article{Herz2025,
  author  = {F. Herz and R. Messina and P. Ben-Abdallah},
  title   = {Broadband directional thermal emission with anisothermal microsources},
  journal = {Physical Review B},
  volume  = {112},
  pages   = {035403},
  year    = {2025},
  doi     = {10.1103/PhysRevB.112.035403}
}

@book{Ozisik1993,
  author    = {M. Necati Öz{\i}\c{s}{\i}k},
  title     = {Heat Conduction},
  edition   = {2th},
  publisher = {John Wiley \& Sons},
  address   = {New York},
  year      = {1993}
}

@book{Whitaker1977,
  author    = {Stephen Whitaker},
  title     = {Fundamental Principles of Heat Transfer},
  publisher = {Pergamon Press},
  address   = {New York},
  year      = {1977}
}

@article{Talukdar2002CDM,
  author  = {Talukdar, Prabal and Mishra, Subhash C.},
  title   = {Analysis of conduction--radiation problem in absorbing, emitting and anisotropically scattering media using the collapsed dimension method},
  journal = {International Journal of Heat and Mass Transfer},
  volume  = {45},
  number  = {10},
  pages   = {2159--2168},
  year    = {2002},
  doi     = {10.1016/S0017-9310(01)00305-2}
}

@article{Venkatraman1997Fluctuation,
  author  = {Venkatraman, Anand and Collins, William D.},
  title   = {The Effect of Temperature Fluctuations on Radiative Heat Transfer in Participating Media},
  journal = {Journal of Quantitative Spectroscopy and Radiative Transfer},
  volume  = {58},
  number  = {1},
  pages   = {75--94},
  year    = {1997},
  doi     = {10.1016/S0022-4073(96)00143-4}
}

@inproceedings{Krebs1996Fluctuations,
  author    = {Krebs, W. and Koch, R. and Ganz, B. and Eigenmann, L. and Wittig, S.},
  title     = {Effect of Temperature and Concentration Fluctuations on Radiative Heat Transfer in Turbulent Flames},
  booktitle = {Proceedings of the Combustion Institute},
  volume    = {26},
  pages     = {2763--2770},
  year      = {1996},
  doi       = {10.1016/S0082-0784(96)80012-4}
}

@article{Saldanha1997,
  author  = {Rog\'erio Martins Saldanha da Gama},
  title   = {The nonlinear conduction/radiation heat transfer phenomenon represented as the limit of a sequence of linear problems},
  journal = {International Communications in Heat and Mass Transfer},
  volume  = {24},
  number  = {1},
  pages   = {119--128},
  year    = {1997},
  publisher = {Elsevier},
  doi     = {10.1016/S0735-1933(96)00111-X}
}

@article{GuentherLee1998,
  author  = {R. B. Guenther and J. W. Lee},
  title   = {Heat conduction with radiating boundary conditions},
  journal = {Journal of Computational and Applied Mathematics},
  volume  = {88},
  pages   = {119--124},
  year    = {1998},
  publisher = {Elsevier},
  doi     = {10.1016/S0377-0427(97)00208-2}
}

@article{LeeChoi2021ThinPlate,
  author  = {Eun-Ho Lee and Woocheol Choi},
  title   = {Asymptotic profile of solutions to the heat equation on thin plate with boundary heating},
  journal = {Applied Mathematics and Computation},
  volume  = {408},
  pages   = {126356},
  year    = {2021},
  publisher = {Elsevier},
  doi     = {10.1016/j.amc.2021.126356}
}

@article{Kim2013EffectiveTemp,
  author  = {Heetae Kim and Myung-Soo Han and David Perello and Minhee Yun},
  title   = {Effective temperature of thermal radiation from non-uniform temperature distributions and nanoparticles},
  journal = {Infrared Physics \& Technology},
  volume  = {60},
  pages   = {7--9},
  year    = {2013},
  publisher = {Elsevier},
  doi     = {10.1016/j.infrared.2013.03.003}
}
\end{document}